\documentclass{article} % For LaTeX2e
\usepackage{iclr2024_conference,times}

\usepackage{amsmath,amsfonts,bm}

\def\eqref#1{equation~\ref{#1}}
\def\1{\bm{1}}

\def\ra{{\textnormal{a}}}

\def\rx{{\textnormal{x}}}

\def\rva{{\mathbf{a}}}

\def\erva{{\textnormal{a}}}

\def\ervx{{\textnormal{x}}}

\def\rmA{{\mathbf{A}}}

\def\vmu{{\bm{\mu}}}
\def\vtheta{{\bm{\theta}}}
\def\va{{\bm{a}}}

\def\ve{{\bm{e}}}

\def\vx{{\bm{x}}}

\def\eva{{a}}

\def\mA{{\bm{A}}}

\def\mH{{\bm{H}}}
\def\mI{{\bm{I}}}
\def\mJ{{\bm{J}}}

\def\mX{{\bm{X}}}

\def\mSigma{{\bm{\Sigma}}}

\DeclareMathAlphabet{\mathsfit}{\encodingdefault}{\sfdefault}{m}{sl}
\SetMathAlphabet{\mathsfit}{bold}{\encodingdefault}{\sfdefault}{bx}{n}
\newcommand{\tens}[1]{\bm{\mathsfit{#1}}}
\def\tA{{\tens{A}}}

\def\tX{{\tens{X}}}

\def\gG{{\mathcal{G}}}

\def\sA{{\mathbb{A}}}
\def\sB{{\mathbb{B}}}

\def\sS{{\mathbb{S}}}

\def\emA{{A}}

\newcommand{\etens}[1]{\mathsfit{#1}}

\def\etA{{\etens{A}}}

\newcommand{\E}{\mathbb{E}}

\newcommand{\R}{\mathbb{R}}

\newcommand{\KL}{D_{\mathrm{KL}}}
\newcommand{\Var}{\mathrm{Var}}

\newcommand{\Cov}{\mathrm{Cov}}
\newcommand{\normltwo}{L^2}
\newcommand{\normlp}{L^p}

\newcommand{\parents}{Pa} % See usage in notation.tex. Chosen to match Daphne's book.

\usepackage{hyperref}
\usepackage{url}
\usepackage{graphicx}
\usepackage{multirow}
\usepackage{bbding}
\usepackage{wasysym}
\usepackage{tikz}
\usepackage{wrapfig}
\usepackage{enumitem}
\usepackage{rotating}
\usepackage{xcolor}
\usepackage{rotating, booktabs}

\usepackage{amssymb}
\usepackage{pifont}
\usepackage[caption=false]{subfig}
\usepackage{ragged2e}
\usepackage{lipsum, caption, booktabs, threeparttable}

\newcommand*\emptycirc[1][1ex]{\tikz\draw (0,0) circle (#1);} 
\newcommand*\halfcirc[1][1ex]{%
  \begin{tikzpicture}
  \draw[fill] (0,0)-- (90:#1) arc (90:270:#1) -- cycle ;
  \draw (0,0) circle (#1);
  \end{tikzpicture}}
\newcommand*\fullcirc[1][1ex]{\tikz\fill (0,0) circle (#1);}

\makeatletter
\def\blfootnote{\xdef\@thefnmark{}\@footnotetext}
\makeatother

\newcommand{\smk}[1]{\textcolor{blue}{(Shoumik: #1)}}

\newcommand{\wwx}[1]{\textcolor{cyan}{(Wenxiao: #1)}}

\title{DRSM: De-Randomized Smoothing on Malware Classifier Providing Certified Robustness}

\author{Shoumik Saha \qquad Wenxiao Wang \qquad Yigitcan Kaya \qquad Soheil Feizi \qquad Tudor Dumitras
\\ \\
\texttt{\{smksaha, wwx, cankaya, sfeizi, tudor\}@umd.edu}\\ \\
Department of Computer Science\\
University of Maryland - College Park
}

\iclrfinalcopy % Uncomment for camera-ready version, but NOT for submission.
\begin{document}

\maketitle

\begin{abstract}
% What is the problem?
Machine Learning (ML) models have been utilized for malware detection for over two decades. Consequently, this ignited an ongoing arms race between malware authors and antivirus systems, compelling researchers to propose defenses for malware-detection models against evasion attacks. 
%
% Why is the problem a problem?
However, most if not all existing defenses against evasion attacks suffer from sizable performance degradation and/or can defend against only specific attacks, which makes them less practical in real-world settings.
%compromise the standard accuracy of the model to some extent that they become impractical in real-life setting. 
%Moreover, some defenses provide robustness against a few selective attacks.
%
% What is our secret sauce? 
In this work, we develop a certified defense, DRSM (De-Randomized Smoothed MalConv), by redesigning the \textit{de-randomized smoothing} technique for the domain of malware detection. 
%The benefit of our proposed \textit{window ablation smoothing} scheme is -- the impact of adversarial byte perturbation is mitigated while the malicious features of a malware file can still be learned. 
Specifically, we propose a \textit{window ablation} scheme to provably limit the impact of adversarial bytes while maximally preserving local structures of the executables.
%
% What did we do?
%We analyze how different window ablations in DRSM can impact its standard and certified accuracy, and compare it to the base classifier.
After showing how DRSM is theoretically robust against attacks with contiguous adversarial bytes, we verify its performance and certified robustness experimentally, where we observe only marginal accuracy drops as the cost of robustness.
%demonstrating its adaptability to various use-case scenarios.
To our knowledge, we are the first to offer certified robustness in the realm of static detection of malware executables. 
%Beyond our theoretical formulations, we extensively evaluate our DRSM models against $9$ different attacks in both white-box and black-box settings.
More surprisingly, through evaluating DRSM against $9$ empirical attacks of different types, we observe that the proposed defense is empirically robust to some extent against a diverse set of attacks, some of which even fall out of the scope of its original threat model.
%
% What are the implications?
%We deliberately include some practical attacks that fall outside of our threat model to showcase the applicability of our proposed method. We then discuss some insights gained from our results.
In addition, we collected $15.5K$ recent benign raw executables from diverse sources, which will be made public as a dataset called PACE (Publicly Accessible Collection(s) of Executables) to alleviate the scarcity of publicly available benign datasets for studying malware detection and provide future research with more representative data of the time.
%to allow future researchers to curate a diverse set of benign programs for experiments that are more representative of the real-world.
%Moreover, to solve the scarcity of publicly available benign datasets in the malware domain, we are releasing our diverse benign dataset containing $15.5K$ programs collected from different sources throughout this work. 
%To facilitate the reproducibility of our findings, we build our framework on top of \texttt{secml-malware} Python library and open-source the implementation. 
%\wwx{not sure if the last sentence is good to have in abstract; probably wouldn't hurt?}
\end{abstract}

\section{Introduction}
\label{sec:intro}

%ML in Malware detection
Machine learning (ML) has started to see more and more adoption in static malware detection, as it also has in many other mission-critical applications. Traditionally, ML models that use static features~\citep{anderson2018ember} require a feature engineering step due to the large size and complex nature of programs. More recently, however, researchers have proposed models like MalConv~\citep{raff2018malware} that can consume whole program simply as raw binary executable to eliminate this step. As expected, there has been a rise in studies showing the adversarial vulnerability of these models in the last few years~\citep{kreuk2018deceiving, lucas2021malware}, resulting in an ongoing arms race.

%Defenses like Adv training and Non-neg suffers
Currently, existing defenses, such as non-negative or monotonic classifier~\citep{fleshman2018non, romeo2018adversarially} and adversarial training~\citep{lucas2023adversarial}, not only introduce sizable drops in standard accuracy but also provide robustness only to specific attacks while still being vulnerable to the rest.

While certified robustness has been studied by many \citep{cohen2019certified, lecuyer2019certified, salman2019provably, levine2020randomized, levine2020robustness}, it remains under-explored in the context of malware detection.
To fill this gap, we redesign the \textit{de-randomized smoothing} scheme, a certified defense originally developed for images~\citep{levine2020randomized}, to detect malware from raw bytes.
With MalConv \citep{raff2018malware} as the base classifier, we use DRSM (De-Randomized Smoothed MalConv) to denote the resulting defense.
To our knowledge, DRSM is the first defense offering \textit{certified robustness} for malware executable detection.
%\can{I'm not entirely sure about this claim, it's too broad. For example, Adversarially Robust Malware Detection Using Monotonic Classification by Incer et al. also provides a provable/verifiable  robustness guarantee with monotonicity, or On Training Robust PDF Malware Classifiers by Chen et al. considers some provable robustness properties for PDF malware classifiers. We need to tune this claim down a bit throughout the paper, it's easy to shoot it down.}

%To our knowledge, we are the first to introduce in malware detection. 

\begin{figure}[tbp]
        \centering
        \includegraphics[width=0.95\linewidth]{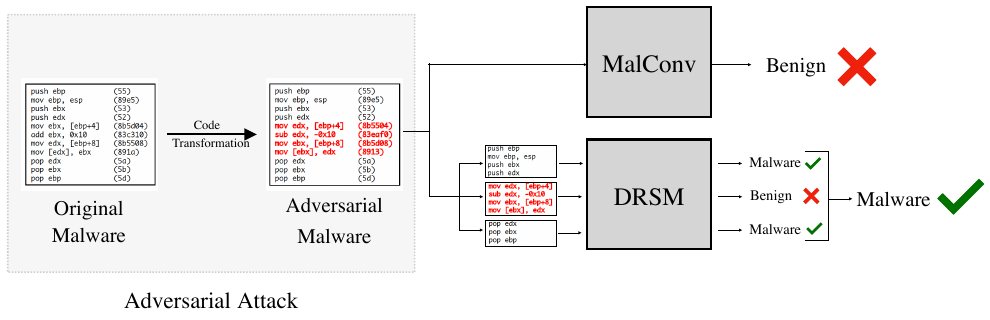}
        \caption{Overview of a prototypical adversarial attack on MalConv and DRSM model. MalConv misclassifies the adversarial malware file as `benign'. Our DSRM creates ablated sequences of the file and makes predictions on each, among which, the majority (\emph{winning}) class is still `malware'.}
        \label{fig:intro_problem}
\end{figure}

%De-randomized smoothing
% In defense of adversarial attack in computer vision (CV), different smoothing schemes have been heavily used for a long time. 
% Recent work ~\citep{levine2020randomized} has improved the \textit{certified robustness} for images by proposing different ablation techniques. Despite of its potential in defending against adversarial attacks, it has still remain untouched by the malware community. 
% We are the first to introduce \textit{de-randomized smoothing} into malware domain and propose \textit{certified robustness}. \wwx{$\Leftarrow$such claim can easily cause trouble;} Using MalConv as the base classifier, we name our proposed model DRSM (De-Randomized Smoothed MalConv). 

%Intrducing DRSM framework
It is challenging for malware domain to utilize de-randomized smoothing scheme due to the inherent difference between image and raw byte file structure. As a solution, we propose a \textit{window ablation} scheme that generates a set of ablated sequences by 
%sliding the window on raw bytes of an input file. 
dividing the input sequence into non-overlapping windows.
%We train the base classifier on these ablated sequences, and consider predictions for each of them. 
For each of these ablated sequences, we train a base classifier keeping the ground truth from original input. At inference, DRSM take the majority of predictions from these base classifiers as its final prediction.
%Aggregating these predictions, DRSM takes the winning class as its final prediction. 
Figure \ref{fig:intro_problem} shows a simplified toy example: An adversarial attack may successfully evaded MalConv model with the presented small changes to the raw executables, but it would still be detected by DRSM if the perturbation could not manipulate sufficient votes.

%Why and how it works?
% Specifically, we propose a \textit{`window ablation'} scheme to address the differences between images and raw byte files, which is an analogue of band ablation in image as illustrated in Figure \ref{fig:drsm_archi}. Intuitively, DRSM should still be able to learn malicious features from its ablated inputs as we are using MalConv model as its base classifier.\wwx{the intuition is not clear} Meanwhile, the perturbed bytes change only a subset of ablated inputs for DRSM model, and therefore its impact on the final prediction can be limited.
% \wwx{It is unclear what these mean without introducing the de-randomized smoothing scheme; either explain de-randomized smoothing earlier or omit these 'intuitions' in intro (the first one is recommended in general) }

%Evaluation
We find that our DRSM ($98.18\%$) can achieve comparable standard accuracy to MalConv ($98.61\%$), and outperforms a prior defense MalConv(NonNeg) ($88.36\%$) by a large margin.
Besides our theoretical formulation for DRSM's certified robustness, we show that it can provide up to $53.97\%$ certified accuracy depending on the attacker's capability. We discuss the performance-robustness trade-offs, and its adaptability upon demand. Moreover, we evaluate the empirical robustness of our DRSM model against $9$ different attacks in both white and black box settings, including attacks outside of the intended threat model of De-Randomized Smoothing. Depending on the attack, even the least robust DRSM model can provide $87.9\% {\sim} 26.5\%$ better robustness than MalConv.

%We experiment with different sizes for our \textit{window ablation} in DRSM models and evaluate their standard and certified accuracy. We find that DRSM ($98.18\%$) can achieve comparable standard accuracy to MalConv model ($98.61\%$), and outperforms MalConv(NonNeg) ($88.36\%$) defense by a large margin. We also provide theoretical formulation of our `certified robustness'  and evaluate it under different perturbation budget. DRSM can provide $53.97\%{\sim}12.2\%$ certified accuracy depending on the ablated window size. We show the trade-off between standard and certified accuracy of DRSM which makes it suitable for different demands. Moreover, we empirically evaluate the robustness of our DRSM model against $9$ different attacks in both white and black box settings, including attacks that fall outside of our threat model. Varying from simple to sophisticated attacks, even the least robust DRSM model can provide $87.9\% {\sim} 26.5\%$ better robustness than MalConv model. 
%\wwx{a bit too detailed; summarizing qualitatively + showing a couple representative results will suffice}

%Dataset and implementation
A practical difficulty in malware research is collecting benign raw executables, due to copyrights and legal restrictions~\cite{anderson2018ember}.
%
%For this work, we collect $15.5K$ fairly recent and diverse benign executables by crawling different sources and compile a dataset of total size $1.1M$ by including datasets from prior studies
Throughout this work, we collect $15.5K$ fairly recent and diverse benign executables from different sources, which can be better representative of the real-world. 
These will be made public as a new dataset, namely PACE (Publicly Accessible Collection(s) of Executables), to alleviate the accessibility issue of benign executables and facilitate future research.
\footnote{Our open-source code and dataset: \url{https://github.com/ShoumikSaha/DRSM/}}
%
%\blfootnote{We implement DSRM on \texttt{secml-malware} Python library with an addition of a minimal number of files, which will be open-sourced to facilitate easy access and extension by future researchers and practitioners.}

Our major contributions include:

\begin{itemize}
    \item A new defense, DRSM (De-Randomized Smoothed MalConv), that pioneers certified robustness in the executable malware domain (Section \ref{sec:de_random_smoothing});
    \item A thorough evaluation of DRSM regarding its performance and certified robustness, which suggests DRSM offers certified robustness with only mild performance degradation (Section \ref{sec:eval});
    \item A thorough evaluation of DRSM regarding its empirical robustness against 9 empirical attacks covering both white-box and black-box settings, which suggests DRSM is empirically robust to some extents against diverse attacks. (Section \ref{sec:emp_robustness})
    \item A collection of $15.5K$ benign binary executables from different sources, which will be made public as a part of our new dataset PACE. (Section \ref{sec:dataset})
    %\item We develop our DRSM framework with minimal addition to the \texttt{secml-malware} Python library which makes it easily adoptable to other different models. We are going to open source our implementation.
    %\item We open source our implementation for DRSM framework, based on \texttt{secml-malware} Python library.
\end{itemize}

\section{Related Work}
\label{sec:related_work}

\textbf{ML in Static Malware Detection.} There have been several studies of how malware executables can be classified using machine learning. As early as 2001 \citet{924286} proposed a data mining technique for malware detection using three different types of static features.
Pioneered by \citet{MalwareImages}, CNN-based techniques for malware detection became popular among security researchers \citep{kalash2018malware, yan2018detecting}. 
Eventually, \citet{raff2018malware} proposed a static classifier, named MalConv, that takes raw byte sequences and detects malware using a convolution neural network. Surprisingly, MalConv is still considered a state-of-the-art for detection with raw byte inputs, potentially attributing to the issue of limited accessibility to benign executables, which we will discuss later in this section. We will use it as base classifiers in this work.

\textbf{Adversarial Attacks and Defenses in Malware Detection.}
Along with the detection research, there has been plenty of research on adversarial attacks on these models. These attacks fall into different categories. For example, attacks proposed by \citet{kolosnjaji2018malware, kreuk2018deceiving, suciu2019exploring} appended and/or injected adversarial bytes in the malware computed by gradient. \citet{demetrio2019explaining, demetrio2021adversarial, nisi2021lost} proposed attacks that modify or extend DOS and Header fields. \citet{demetrio2021functionality} extracted payloads from benign files to be appended and injected into malware files. Recent work by \citet{lucas2021malware} used two types of code transformation to generate adversarial samples.
%\textbf{Prior Defenses for Adversarial Attacks.} 
For defenses, \citet{fleshman2018non} proposed a defense, MalConv (NonNeg), by constraining weights in the last layer of MalConv to be non-negative. However, this model achieves low accuracy of $88.36\%$, and has been shown to be as vulnerable as MalConv in some cases \citep{wang2023mpass, wang2022black, ceschin2019shallow}. 
Another defense strategy, adversarial training cannot guarantee defense against attacks other than the one used during training, which limits its usage: \citet{lucas2023adversarial} showed training it on Kreuk-0.01 degraded the true positive rates to $84.4\% \sim 90.1\%$.
%However, recent work \citep{lucas2023adversarial} showed that adversarial training on one attack cannot guarantee defense against other attacks which makes it inadequate for real-world use. Also, they showed training it on Kreuk-0.01 dropped the TPR to $84.4\% \sim 90.1\%$. 
Notably, where variants of randomized smoothing schemes have been proposed for vision domains \citep{cohen2019certified, lecuyer2019certified, salman2019provably, levine2020randomized, levine2020robustness}, they remain under-explored in the context of malware detection.

\textbf{Limited Accessibility to Benign Executables.}
Though there have been a large amount of work on malware detection, most of the works were done using private or enterprise dataset with restrictive access. 
%And most of them could not publish the benign raw executables due to copyright issue \citep{lucas2021malware}. 
Prior works \citep{anderson2018ember, yang2021bodmas, downing2021deepreflect} explain the copyright issue and only published the feature vector of benign files (see Table \ref{table:other_dataset}). This impose many constraints to the advancement of malware detection techniques, especially to have a complete model that requires raw executables as inputs.
%To solve this issue, we are going to publish our benign dataset following common standards.

\section{Background and Notations}
\label{sec:backgorund}

We denote the set of all bytes of a file as $X \subseteq [0, N-1]$, where $ N = 256$. 
A binary file is a sequence of k bytes $x = (x_1, x_2, x_3, ... x_k)$, where $ x_i \in X$ for all $1 \leq i \leq k$. Note that the length $k$ varies for different files, thus $k$ is not fixed. However, the input vector fed into the network has to be of a fixed dimension. So, the common approach is to -- pad zeros at the end of $x$ if $k < D$, or extract the first $D$ bytes from $x$, to fix the input length to $D$.

\subsection{Base Classifier}
\label{sec:base_classifier}

In this work, we will be using the state-of-the-art static malware detection model to this date, named MalConv \citep{raff2018malware}, as our base classifier.
While there are other models like Ember, GBDT \citep{anderson2018ember} for malware detection, note that -- these models can work only on feature vectors, and our work focuses on raw binary executables files.
 Let us represent the MalConv model (see Figure \ref{fig:malconv_model}) as $F_{\theta} : X \rightarrow [0,1]$ with a set of parameters $\theta$ that it learns through training. If the output of $F_{\theta}(X)$ is greater than $0.5$ then the prediction is considered as 1 or malicious, and vice versa. We set the input length as $2$MB following the original paper.

%\smk{we can take this para to appendix if needed}\\
MalConv takes in each byte $x_i$ from file $X$ and then passes it to an embedding layer with an embedding matrix $Z \in \mathbb{R}^{D \times 8}$, which generates an embedding vector $z_i = \phi (x_i)$ of 8 elements. This vector is then passed through two convolution layers, using ReLU and sigmoid activation functions. These activation outputs are combined through a gating which performs an element-wise multiplication to mitigate the vanishing gradient problem. The output is then fed into a temporal max pooling layer, followed by a fully connected layer. Finally, a softmax layer calculates the probability of $X$ being a malware or benign file.

\subsection{Threat Model}
\label{sec:threat_model}

Unless otherwise specified, we assume that the attacker has the full knowledge of the victim, including architectures and model parameters. This is typically referred to as the white-box setting. The white-box setting considers potentially strong attackers, which is desired when assessing defenses.

%From a defender's perspective, such a white-box manner is realistic and conservative since -- `security through obscurity' is not a good approach, and it is possible to deploy transfer attack or query-based attack even in the black-box setting \citep{papernot2016transferability, demetrio2021functionality}.

In the primary threat model that we consider when developing our defense, the attacker can modify or add any bytes in a contiguous portion of the input sample in test time to evade the model. So, the goal of the attacker is to generate a perturbation $\delta$ that creates the adversarial malware $x^{'} = x + \delta$, for which $F_{\theta}(x^{'}) < 0.5$, i.e., the classifier predicts it as a benign file. Here, the attacker knows the classifier model $F$ and its parameters $\theta$, and can modify the original malware file $x$. However, finding the perturbation $\delta$ in a binary file is more challenging than vision due to its inherited file structures. For any arbitrary change in a malware file, the file can lose its semantics, i.e. malicious functionality, in the worst case, the file can get corrupted. 
%We assume that the attacker is powerful enough to find such perturbation $\delta$.

Even after such challenges in binary modification, prior attacks have been successful by adding contiguous adversarial bytes in one \citep{kreuk2018deceiving, demetrio2021adversarial} or multiple locations \cite{suciu2019exploring, demetrio2021functionality}, or modifying bytes at specific locations\citep{demetrio2019explaining, nisi2021lost}, to evade a model. 
In this work, we consider not only the former ones, which fall directly within our primary threat model but also the latter ones which do not. 
In addition, we also consider recent, more sophisticated attacks \citep{lucas2021malware, lucas2023adversarial} where the attacker has the power to disassemble malware and apply different code transformations at any place in the file. For coherence, we defer the details about these attacks to Section \ref{sec:emp_robustness}, where we evaluate the empirical robustness of our defenses against them.
%We emphasize that -- though such attacks fall outside of our threat model, we consider them to show that -- the empirical robustness of our model surpasses the theoretical one (see Section \ref{sec:emp_robustness}).

\section{A New Publicly Available Dataset---PACE}
\label{sec:dataset}

%\citep{cao2020benign} -- This paper shows that 29.5\% of popular features are generated by benign samples. Thus, it is easier for a machine learning model to learn these benign features and make a decision based on the existence of such features. So, a diverse set of benign samples should cover more features and can make a model more robust which led us to create a diverse dataset to train and evaluate our model.

%\citep{anderson2018ember} -- EMEBR paper discusses why it is tough to collect benign samples due to the copyright law and legal restrictions. And they did not publish the raw executables of the benign files.

%\citep{downing2021deepreflect} -- This paper discusses the importance of diverse benign samples, and they crawled .CNET and MSI to collect benign executables whereas we went further by doing that on other sources (sourceforge, NET, softonic, etc.) including .CNET.

%\citep{yang2021cade, jordaney2017transcend, barbero2022transcending} -- papers on concept drift in malware detection.

Like other domains, malware detection suffers from concept drift too. Previously, \citet{yang2021cade, jordaney2017transcend, barbero2022transcending} demonstrated how concept drift can have a disastrous impact on ML-based malware detection. Therefore, we used $3$ datasets from different times in this work (Table \ref{table:dataset}). However, in the malware domain, having a large dataset to train a machine learning (ML) model may not be enough as maintaining diversity and recency is also crucial \citep{cao2020benign, downing2021deepreflect}.
We found that models trained without diverse benign samples can have a very high false positive rate (see details in Appendix \ref{app:pace_degrade}).
%Otherwise, ML model can get inclined to learn spurious features related to specific signatures in benign binaries (see Appendix \ref{app:pace_degrade}). 
%For example, \citet{cao2020benign, downing2021deepreflect} emphasized the importance of diversity in benign dataset. 

%\subsection{PACE Dataset}
Despite the importance of diverse benign samples, unfortunately, most prior works (\citet{anderson2018ember, downing2021deepreflect}) could not publish raw executables of benign files due to copyright and legal restrictions.
For this work, we crawled popular free websites, e.g., SourceForge, CNET, Net, Softonic, etc., to collect a diverse benign dataset of size 15.5K (Table \ref{table:benign_dataset}), naming \textbf{PACE} (Publicly Accessible Collection(s) of Executables).
We collected the malware from \href{https://virusshare.com/}{VirusShare} at the same time (August 2022) as benign files.
Following the common practice and guidelines, we are publishing the URLs along with the MD5 hash for each raw benign file in our dataset (see Appendix \ref{app:dataset} for more details). We hope this will help researchers to recreate the dataset easily and experiment with a better representative of real-world settings in the future.\footnote{PACE malware samples will also be provided upon request.}

%Besides diversity, we also consider the temporality of datasets in this work. Prior works \citep{yang2021cade, jordaney2017transcend, barbero2022transcending} demonstrated how concept drift can have disastrous impact in malware detection. Therefore, we used $3$ datasets from different times in this work (Table \ref{table:dataset}) for training and made sure that the test set comes from the latest one during evaluation. We collected the malware files that was added to VirusShare \footnote{\url{https://virusshare.com/}} at the same time when we collected the benign files (August, 2022).  

\begin{table}[!htb]
    
    \begin{minipage}{.6\linewidth}
      \caption{Datasets used in this work with collection time, size, and public availability of raw executables}
      \centering
        \resizebox{\linewidth}{!}{
        \begin{tabular}{ llcccc }
        \hline
          Dataset & Collection  & \multicolumn{3}{c}{Number of Binaries} & Public  \\\cline{3-5}
          Name & Time & Malware & Benign & Total & Availability \\
        \hline 
        Ember  & 2017 & 400K & 400K & 800K & \XSolidBrush \\
        VTFeed & 2020 & 139K & 139K & 278K & \XSolidBrush\\
        PACE (Our) & 2022 & 15.5K & 15.5K & 31K & \CheckmarkBold\\
         \hline 
        Total & & 554.5K & 554.5K & 1.1M\\
          \hline
        \end{tabular}
        }
    
    \label{table:dataset}
    \end{minipage}%
    \begin{minipage}{.4\linewidth}
      \centering
        \caption{PACE (Benign) Dataset}
        \resizebox{\linewidth}{!}{
        \begin{tabular}{ lc }
        \hline
          Source  & Number of Binaries \\
         \hline
        SourceForge  & 7,865 \\
        CNET  & 3,661 \\
        Net  & 2,534 \\
        Softonic  & 1,152 \\
        DikeDataset  & 1,082 \\
        Netwindows  & 185 \\
        Manually Obtained   & \multirow{2}{*}{89} \\
        from Windows OS & \\
         \hline
        Total & 15,568\\
          \hline
        \end{tabular}
        }
    
    \label{table:benign_dataset}
    \end{minipage} 
    %\caption{Global caption}
\end{table}

We used a MalConv model pre-trained on Ember \citep{anderson2018ember} dataset provided by the \href{https://en.wikipedia.org/wiki/Endgame,_Inc.}{Endgame Inc.} 
Then we used this model to re-train the MalConv, MalConv (NonNeg), and our DRSM models on both VTFeed  and PACE (our) dataset.\footnote{The authors of \citep{lucas2021malware} assisted in training models on VTFeed, which we could not have done by ourselves since VTFeed is not publicly accessible} We split our dataset into 70:15:15 ratios for train, validation, and test sets, respectively. During evaluation, we made sure that test samples came from the latest dataset (PACE) only.
%We trained and evaluated our models using multiple NVIDIA RTX A4000 and 2 RTX A5000 gpus. 
For more details about model implementation, see Appendix \ref{app:model_impl}.

\iffalse

\begin{table}[h]
\centering
\begin{tabular}{ llcccc }
    \hline
      Dataset & Collection  & \multicolumn{3}{c}{Number of Binaries} & Public  \\\cline{3-5}
      Name & Time & Malware & Benign & Total & Availability \\
    \hline 
    Ember  & 2017 & 400K & 400K & 800K & \XSolidBrush \\
    VTFeed & 2020 & 139K & 139K & 278K & \XSolidBrush\\
    Our Dataset & 2022 & 15.5K & 15.5K & 31K & \CheckmarkBold\\
     \hline 
    Total & & 554.5K & 554.5K & 1.1M\\
      \hline
\end{tabular}
\caption{Datasets used for our work}
\label{table:dataset}
\end{table}

\begin{table}[h]
  %\centering
\begin{center}
\begin{tabular}{ ll }
    \hline
      Source  & Number of Binaries \\
     \hline
    SourceForge  & 7,865 \\
    CNET  & 3,661 \\
    Net  & 2,534 \\
    Softonic  & 1,152 \\
    DikeDataset  & 1,082 \\
    Netwindows  & 185 \\
    Manually Obtained from Windows OS  & 89 \\
     \hline
    Total & 15,568\\
      \hline
\end{tabular}
\end{center}
\caption{Our Published Benign Dataset}
\label{table:benign_dataset}
\end{table}

\fi

\section{DRSM: De-Randomized Smoothing on Malware Classifier}
\label{sec:de_random_smoothing}

%For certified defense against the adversarial attack, smoothed vision models have long been used in the vision domain. However, such defense method is still untouched by malware community. 
%The intuition behind incorporating the `de-randomized smoothing' technique with raw binary file is -- when the attacker generates a byte perturbation attacking the base classifier, it is not necessary that those bytes will have the same impact on all ablated inputs of the smoothed classifier. Besides, the bytes that contain the malicious features remain preserved in the ablated versions, and hence, the smoothed classifier can still learn the malicious features. 

Since the malware detection problem cannot be directly mapped to typical vision problems, we had to redesign the `de-randomized smoothing' scheme to make it compatible. Unlike images, our input samples are one-dimensional sequences of bytes, which makes the common vision-oriented ablation techniques, e.g., adding noise, masking pixels, block ablations, etc., infeasible. Additionally, even a random byte change in a file may cause a behavior change or prevent the sample from executing. 

\begin{figure}[h]
        \centering
        \includegraphics[width=1.0\linewidth]{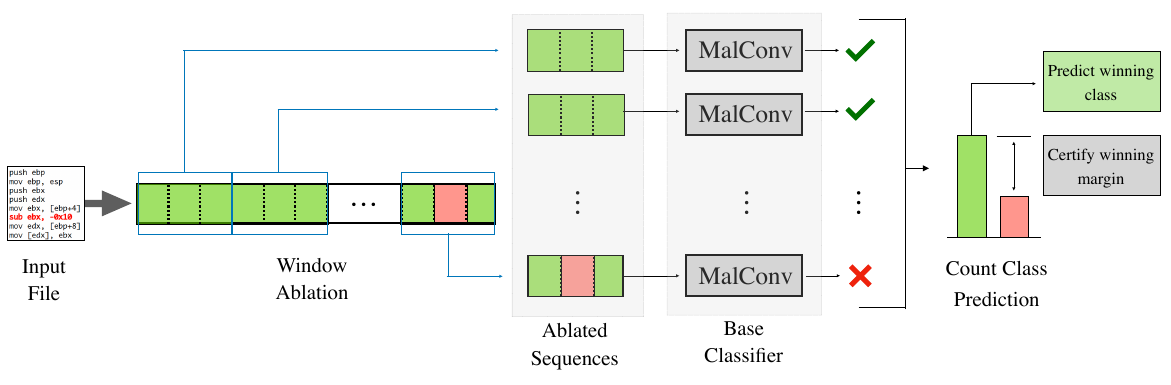}
        \caption{DRSM (De-Randomized Smoothed MalConv) model framework}
        \label{fig:drsm_archi}
\end{figure}

So, we introduce the \textit{`window ablation'} strategy which involves segmenting the input sample into multiple contiguous sequences of equal size. If the input length of the base classifier is $L$, and the size of the ablated window is $w$, then there will be $\lceil \frac{L}{w} \rceil$ ablated sequences of length $w$ resulting in the ablated sequence set $S(x)$. So, even if an attacker generates a byte perturbation of size $p$, it can modify at most $ \Delta = \lceil \frac{p}{w} \rceil + 1$ ablated sequences ($+1$ when a perturbation overlaps $2$ windows). Since a perturbation can only influence a limited number of ablated sequences, it cannot directly change the decision of the smoothed-classifier model -- which was our prior motivation to integrate this technique. A visual representation of our strategy is provided in Figure \ref{fig:drsm_archi}. 

The goal of the defender is to -- using $F_\theta$ as the base classifier, find a \textit{de-randomized smoothed model} $G_\theta$  that can detect any adversarial malware $x^{'}$ generated using a perturbation $\delta$. $G_\theta$ takes in each sequence $s$ from the ablated sequence set $S(x)$, and returns the most frequent predicted class. Specifically, for an input file $x$, ablated sequence set $S(x)$, and base classifier $F_\theta$, the \textit{de-randomized smoothed model} $G_\theta$ can be defined as:
$$ G_{\theta}(x) = \arg \max_{c} n_c(x) $$
where, 
$$ n_c(x) = \sum_{x^{'} \in S(x)} I\{F_{\theta}(x^{'})=c\} $$ denotes the number of ablated sequences that were predicted as class $c$. The percentage of files that are correctly classified by the \textit{de-randomized smoothed model} $G_\theta$ is the \textbf{`standard accuracy'}. 

We say the classifier $G_\theta$ \textit{certifiably robust} on an ablated sequence set if the number of predictions for the correct class exceeds the incorrect one by a `large margin' (dictated by byte size of perturbation). This `large margin' puts a lower bound on attacker's success in altering predictions of the classifier $G_\theta$ since a perturbation $\delta$ of size $p$ can, at most, impact  $\Delta = \lceil \frac{p}{w} \rceil + 1$ ablated sequences.

Mathematically, the \textit{de-randomized smoothed model} $G_{\theta}$ is `certifiably robust' on input $x$ for predicting class $c$ if:
$$ n_c(x) > max_{c \neq c^{'}} n_{c^{'}}(x) + 2 \Delta $$
Since our problem is a binary classification problem, this can be rewritten as:
\begin{equation} \label{eqn:cond_certtify}
\begin{split}
 n_m(x) > n_b(x) + 2 \Delta \ \ \ \ \ & \text{; if $true{-}label(x) = malware$}\\
 n_b(x) > n_m(x) + 2 \Delta \ \ \ \ \ & \text{; if $true{-}label(x) = benign$}
\end{split}
\end{equation}

%$$ n_m(x) > n_b(x) + 2 \Delta $$
where, $n_m(x)$ and $n_b(x)$ are the number of ablated sequences predicted as malware and benign by the \textit{de-randomized smoothed model} $G_{\theta}$, respectively. The percentage of file that holds the inequality \ref{eqn:cond_certtify} for $G_{\theta}$ is the \textbf{`certified accuracy'}.

%\smk{Should I omit the following paragraph here? and introduce DRSM in evaluation section?}

%As the base classifier, we used MalConv model, and so named the de-randomized smoothed model  as DRSM (De-Randomized Smoothed MalConv). 
For simplicity, we will use DRSM-n to denote DRSM with the number of ablated sequences $|S(x)| = \text{n}$, e.g. DRSM-4 means $4$ ablated sequences on input $x$ will be generated for DRSM.

\section{Certified Robustness Evaluation}
\label{sec:evaluation}

\label{sec:eval}

\subsection{Standard Accuracy}
\label{sec:accuracy}

For evaluation, we compare our DRSM models with 
%its base classifier 
MalConv\citep{raff2018malware} which is still one of the state-of-the-art models for static malware detection. 
%While there are other models like Ember\citep{anderson2018ember}, note that -- this work focuses on raw binary executables files, and such models can work only on feature vectors. 
Moreover, we consider the non-negative weight constraint variant of MalConv which was proposed as a defense against adversarial attack in prior work \citep{fleshman2018non}. We train and evaluate these models on the same train and test set (Section \ref{sec:dataset}).

\begin{table}[h]
  %\centering
  \caption{Standard and Certified Accuracy of Models. MalConv and MalConv(NonNeg) cannot provide certified accuracy}
\begin{center}
\resizebox{\linewidth}{!}{
\begin{tabular}{ lccc|ccc }
    \hline
      \multirow{2}{*}{Model}  & \multicolumn{3}{c}{Standard Accuracy (in \%) $\uparrow$} & \multicolumn{3}{c}{Certified Accuracy [$\Delta=2$] (in \%) $\uparrow$}\\\cline{2-4}\cline{5-7}
             & Train-set & Validation-set & Test-set & Train-set & Validation-set & Test-set \\
     \hline
    MalConv  & $99.73$ & $98.87$ & $98.61$ & $-$ & $-$ & $-$\\
    MalConv(NonNeg) & $88.56$ & $87.56$ & $88.36$ & $-$ & $-$ & $-$\\
    DRSM-4   & $99.49$ & $98.12$ & $98.18$ & $14.74$ & $7.84$ & $12.2$\\
    DRSM-8   & $99.67$ & $97.88$ & $97.79$ & $52.74$ & $43.9$ & $40.85$\\
    DRSM-12  & $96.07$ & $95.58$ & $95.88$ & $45.77$ & $44.43$ & $46.21$\\
    DRSM-16  & $94.29$ & $93.00$ & $93.3$ & $59.1$ & $50.52$ & $49.17$\\
    DRSM-20  & $91.17$ & $91.05$ & $91.15$ & $51.64$ & $51.92$ & $52.68$\\
    DRSM-24  & $90.22$ & $89.80$ & $90.24$ & $54.19$ & $54.88$ & $53.97$\\
     \hline
\end{tabular}
}
\end{center}

\label{table:std_acc}
\end{table}

For DRSM-n, we choose $n= \{ 4,8,12,16,20,24 \}$ for our experiments and show the standard accuracy on the left side of the Table \ref{table:std_acc}. 
%Notably, `standard accuracy' is the percentage of files that are correctly classified by a model. 
Recall that -- for DRSM-n, a file is correctly classified if the winning class from majority voting matches the true label for that file (Section \ref{sec:de_random_smoothing}). For ties, we consider `malware' as the winning class. From the Table \ref{table:std_acc}, we can see that -- DRSM-4 ($98.18\%$) and DRSM-8 ($97.79\%$) can achieve comparable accuracy to the MalConv model ($98.61\%$). However, increasing the $n$ has a negative impact on the standard accuracy. For example, DSRM-20 and DSRM-24 achieve $91.15\%$ and $90.24\%$ standard accuracy, respectively. We investigate and find that -- with more ablations (smaller window), the probability of one window containing enough malicious features to make a stable prediction becomes less.
%and thus, the model starts struggling to learn those features. 
On the other hand, the MalConv (NonNeg) model has a lower accuracy, which is consistent with the results by \citet{fleshman2018non}.

\subsection{Certified Accuracy}

Besides standard accuracy, we also evaluate the certified accuracy for DRSM-n models. Recall that -- `certified accuracy' is the percentage of files for which the inequality \ref{eqn:cond_certtify} holds true for DRSM-n models. In short, it denotes the lower bound of model performance even when the attacker can perturb bytes in $\Delta$ number of ablated windows and alter predictions for all of them. 
So, we run experiments on DRSM-n models by varying the $\Delta$ in equation \ref{eqn:cond_certtify}, i.e., perturbation budget for the attacker.
To maintain consistency between standard and certified accuracy, we take `malware' as the winning class for ties by tweaking the first inequality in \ref{eqn:cond_certtify} to $n_m(x) \geq n_b(x) + 2 \Delta$.
%Since in case of equal votes, we take the malware as winning class for standard accuracy, we follow the same rule for certified accuracy by tweaking the first inequality in \ref{eqn:cond_certtify} to $n_m(x) \geq n_b(x) + 2 \Delta$. 

\begin{figure}[tbp]
        \centering
        \includegraphics[width=1.0\linewidth]{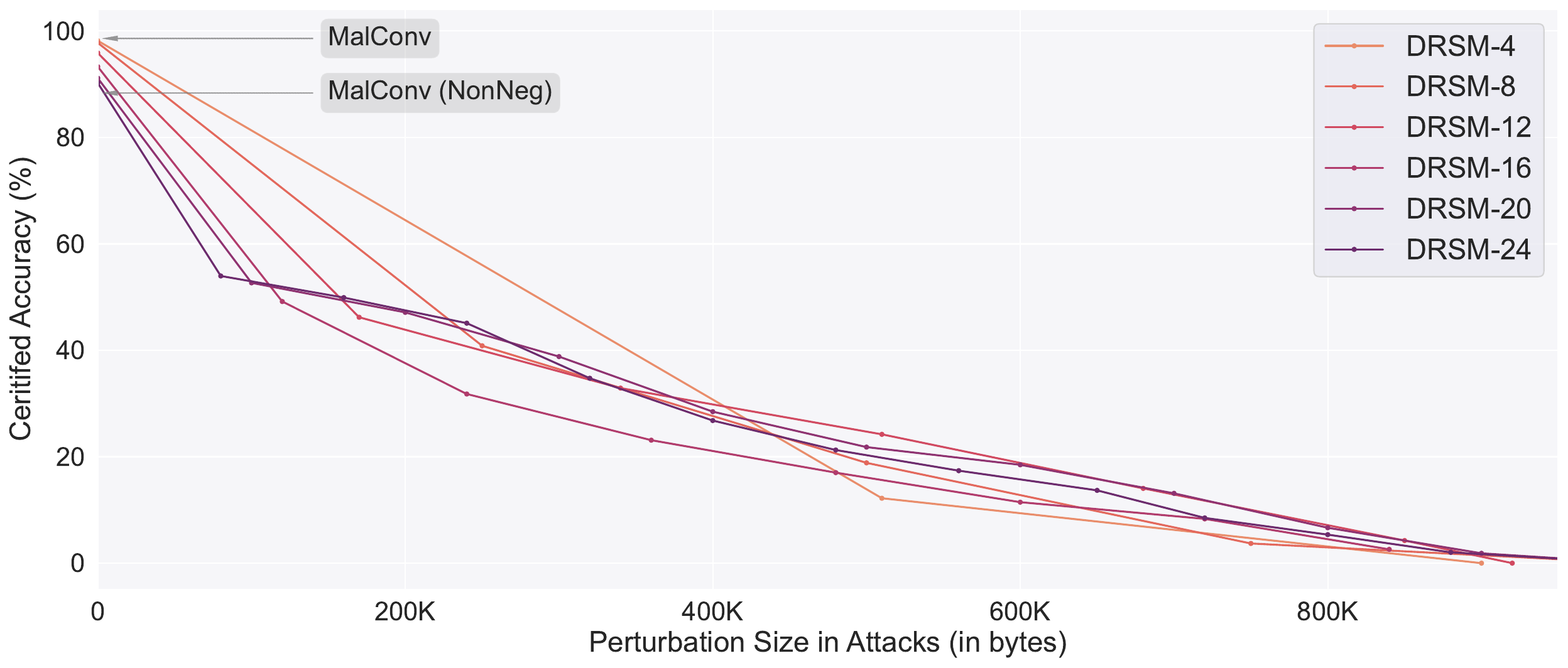}
        \caption{Certified Accuracy (\%) of DRSM-n models for different perturbation budgets (Test-set). While MalConv and MalConv (NonNeg) are not certifiably robust, their standard accuracy is highlighted for references.}
        \label{fig:cert_acc_test_plot}
\end{figure}

%Notably, the range of $\Delta$ starts from $2$ to the point where the inequality \ref{eqn:cond_certtify} becomes impossible to be true, i.e., $\Delta \leq \frac{n}{2}$. 
Notably, $\Delta \in \{ 2, 3, ... , \frac{n}{2} \}$. The range starts from $2$, because any perturbation smaller than the window size can overlap with at most $2$ ablated sequences, and goes up to $\frac{n}{2}$, because the inequality \ref{eqn:cond_certtify} will never hold beyond this point.
The right side of Table \ref{table:std_acc} shows the certified accuracy of DRSM-n models for $\Delta=2$.
In Figure \ref{fig:cert_acc_test_plot}, we show the result of certified accuracy on the test set for all values of $\Delta$, i.e., perturbation budget for the attacker (x-axis). See Tables \ref{table:cert_acc_all} and \ref{table:cert_acc_range} in Appendix \ref{app:results} for more details. We emphasize that even with small $\Delta=2 (=\lceil \frac{255K}{256K} \rceil + 1)$, an attacker can change up to $255K$ bytes for DRSM-8, and yet the model maintains $40.85\%$ certified accuracy.

%Right side of Table \ref{table:std_acc} shows the ranges of certified accuracy for different models. The range starts from $\Delta = 1$ to the point where the inequality \ref{eqn:cond_certtify} becomes impossible to be true, i.e., $\Delta < \frac{n}{2}$. For DRSM-4, there is only one value because $\Delta=\{1\}$, and note that -- even with such small $\Delta$, an attacker can change up to $511K$ bytes and perturb at most $2$ window ablations (when a perturbation overlaps multiple windows) in DRSM-4. For example, in DRSM-16, if the attackers have a perturbation budget up to $200K$ bytes, they can perturb  at most $\Delta = \lceil \frac{p}{w} \rceil + 1 = \lceil \frac{200K}{128K} \rceil + 1 = 3$ window ablations, i.e., alter at most $3$ predictions, and still the model would provide $31.79\%$ certified accuracy. Figure \ref{fig:cert_acc_test_plot} shows the change in certified accuracy for each model with varying the perturbation budget. 

By analyzing Table \ref{table:std_acc}, we can see that $n$ has a positive and negative correlation with certified and standard accuracy, respectively. While DRSM-24 provides the highest certified accuracy ($53.97\%$), it has the lowest standard accuracy ($90.24\%$) among all DRSM-n models. In contrast, DRSM-4 provides the highest standard accuracy ($98.18\%$) with $12.2\%$ certified accuracy. 
%Though the performance seems like a trade-off, models like DRSM-8, DRSM-16 achieves high certified robustness along with considerably good standard accuracy, while prior defense like MalConv (NonNeg) achieves low standard accuracy like $88.36\%$. 
This observation may suggest a performance trade-off.
However, it's worth highlighting that models like DRSM-8 and DRSM-16 strike a balance, delivering robust certified performance alongside commendable standard accuracy, while prior defense MalConv (NonNeg) achieves lower standard accuracy $88.36\%$.
We also want to emphasize that -- perturbing even $200$KB in a $2$MB file ($=10\%$)  is very challenging in a malware file, and yet our DRSM-n models can provide $37\% {\sim} 64\%$ certified accuracy for such perturbation (from Figure \ref{fig:cert_acc_test_plot}). Remember that -- this accuracy reports the theoretical lower bound and in practice, our DRSM-n models provide even higher robustness (shown in Section \ref{sec:emp_robustness}).

\section{Empirical Robustness Evaluation}
\label{sec:emp_robustness}

Beyond theoretical robustness, we also evaluate the empirical robustness of our DRSM-n models. Recall from Section \ref{sec:threat_model} that -- in our threat model (any de-randomized smoothing scheme), attackers can add, or modify bytes in a contiguous portion of a malware file, to get it misclassified as a benign one. However, in real-life settings, attackers can be more capable and can deploy complex attacks where they can find multiple contiguous blocks to perturb. 
%Moreover, there are very recent works where attacker can even apply different code transformations guided by the model to modify bytes at many different places all over the file. 

%For a security-sensitive application like malware detection, a careful evaluation against different attacks is necessary, irrespective of the threat model. 
%Unfortunately, most prior defenses have neglected this. 
In this work, we consider $9$ different attacks in both white and black box settings and categorize them into $3$ types based on their alignment with our threat model. \textbf{Fully Aligned: }if an attack perturbs bytes in one contiguous block; \textbf{Partially Aligned: }if an attack perturbs bytes in multiple different contiguous blocks; \textbf{Not Aligned: }if an attack applies code transformation and changes bytes all over the file (not limited to any contiguous block). Table \ref{table:attacks} shows the list of attacks that have been considered in this work along with their type, settings, and short description. For more details about individual attacks and their implementation, see Appendix \ref{app:attack_app}.

\iffalse
\begin{itemize}[leftmargin=*]
    \item If an attack perturbs bytes in one contiguous block, then it is \textbf{fully inside} our threat model. Example - FGSM Append, DOS Extension, DOS Modification (Partial, Full), etc.
    \item If an attack perturbs bytes in multiple different contiguous blocks, then it is \textbf{partially inside} our threat model. This threat model subsumes the previous one. Example - Slack Append, Header Field Modification, GAMMA, etc.
    \item If an attack applies code transformation and changes bytes all over the file (not limited to any contiguous block), then it is \textbf{outside} of our threat model. Example - Disp, IPR, etc.
\end{itemize}
\fi

%Since attacks in the white-box setting is more powerful (or capable) than the balck-box one, our major focus was on the former one. However, we included some black-box attacks as well. 

\begin{table}[h]
  %\centering
  \caption{Attacks evaluated. \emptycirc[0.75ex] - Fully Aligned, \halfcirc[0.75ex] - Partially Aligned and \fullcirc[0.75ex] - Not Aligned describe the alignment of the attacks to our primary threat model (see Section \ref{sec:threat_model}).}
\begin{center}
\resizebox{\linewidth}{!}{
\begin{tabular}{ lcccp{6cm} }
    \hline
      \multirow{2}{*}{Attack}  & Threat & \multicolumn{2}{c}{Settings} & Short Description \\\cline{3-4}
             &   Model   &    White-box & Black-box & \\
     \toprule 
    \multirow{2}{*}{\shortstack[l]{FGSM Append \\ \citep{kreuk2018deceiving}}} & \multirow{2}{*}{\emptycirc[1ex]} & \multirow{2}{*}{\CheckmarkBold} &  & Appends random bytes at the end of the file generated by FGSM \\ \midrule
    
    \multirow{2}{*}{\shortstack[l]{Slack Append \\ \citep{suciu2019exploring}}} & \multirow{2}{*}{\halfcirc[1ex]} & \multirow{2}{*}{\CheckmarkBold} &  & Injects non-functional bytes in slack regions generated by FGSM \\ \midrule
    
    \multirow{2}{*}{\shortstack[l]{DOS Extension \\ \citep{demetrio2021adversarial}}} & \multirow{2}{*}{\emptycirc[1ex]} & \multirow{2}{*}{\CheckmarkBold} &  & Extends the DOS header and injects adversarial noise\\ \midrule
    
    \multirow{2}{*}{\shortstack[l]{DOS Modification (Partial) \\ \citep{demetrio2019explaining}}} & \multirow{2}{*}{\emptycirc[1ex]} & \multirow{2}{*}{\CheckmarkBold} & \multirow{2}{*}{\CheckmarkBold} & Puts adversarial noise in between of \texttt{MZ} and offset \texttt{0x3c} in the DOS header\\ \midrule
    
    \multirow{2}{*}{\shortstack[l]{DOS Modification (Full) \\ \citep{demetrio2021adversarial}}} & \multirow{2}{*}{\emptycirc[1ex]} & \multirow{2}{*}{\CheckmarkBold} & \multirow{2}{*}{\CheckmarkBold} & Modifies every byte in the DOS header without corrupting the file\\ \midrule
    
    \multirow{2}{*}{\shortstack[l]{Header Field Modification \\ \citep{nisi2021lost}}} & \multirow{2}{*}{\halfcirc[1ex]} & \multirow{2}{*}{\CheckmarkBold} & \multirow{2}{*}{\CheckmarkBold} & Modifies fields in PE header\\ \\ \midrule
    
    \multirow{2}{*}{\shortstack[l]{Disp \\ \citep{lucas2021malware}}} & \multirow{2}{*}{\fullcirc[1ex]} & \multirow{2}{*}{\CheckmarkBold} &  & Displaces code instructions using \texttt{jmp} and semantic \texttt{nop} \\ \midrule
    
    \multirow{2}{*}{\shortstack[l]{IPR \\ \citep{lucas2021malware}}} & \multirow{2}{*}{\fullcirc[1ex]} & \multirow{2}{*}{\CheckmarkBold} &  & Replaces instructions in multiple ways (equiv. replace, register reassign, reorder, etc.) without altering functionalities\\ \midrule
    
    \multirow{2}{*}{\shortstack[l]{GAMMA \\ \citep{demetrio2021functionality}}} & \multirow{2}{*}{\halfcirc[1ex]} &  & \multirow{2}{*}{\CheckmarkBold} & Extracts payloads from benign programs and injects them in malware\\
      \midrule
\end{tabular}
}
\end{center}

\label{table:attacks}
\end{table}

%\subsection{Attack Evaluation}
To evaluate the attacks against MalConv, MalConv (NonNeg) and DRSM-n models, we randomly sampled $200$ malware from the test-set of our dataset that are correctly classified by the model before attack. Let us call this subset of malware the `attack set'. 
%We could use higher number of samples, but some attacks take very long time to generate an adversarial sample and for a fair comparison, we wanted to use the same set for every attack.
We call an attack `successful' if the attack can generate a functional adversarial malware that can change the model's prediction from `malware' to `benign'.
Even though the majority voting in DRSM-n is not differentiable, it can still be attacked by targeting its base classifiers. Correspondingly, whenever necessary, we generate adversarial malware from the `attack set' by differentiating through the base classifier.
%Note that, for non-negative weight constraint, it is not possible to directly attack MalConv (NonNeg) model.

%\citep{wang2022black, ceschin2019shallow, lucas2021malware, wang2023mpass} -- shows that Non-Neg MalConv is vulnerable

\begin{figure}[tbp]
    \subfloat{
    \includegraphics[clip,width=1.0\linewidth]{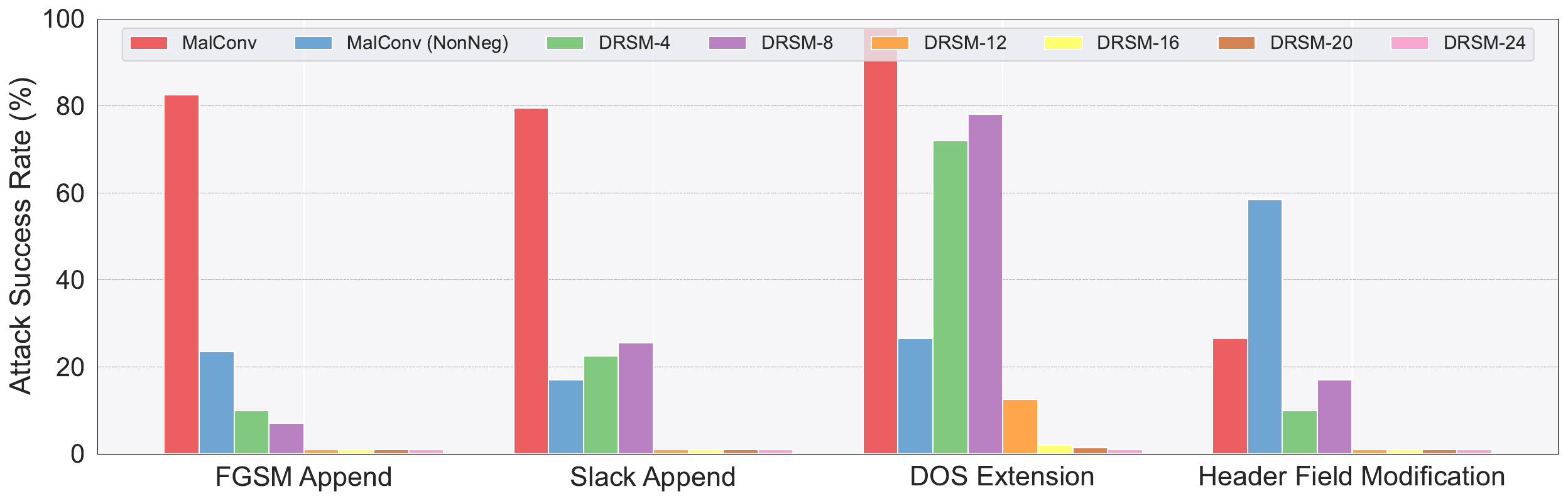}%
    }\\[-3ex]
    \subfloat{
    \includegraphics[clip,width=1.0\linewidth]{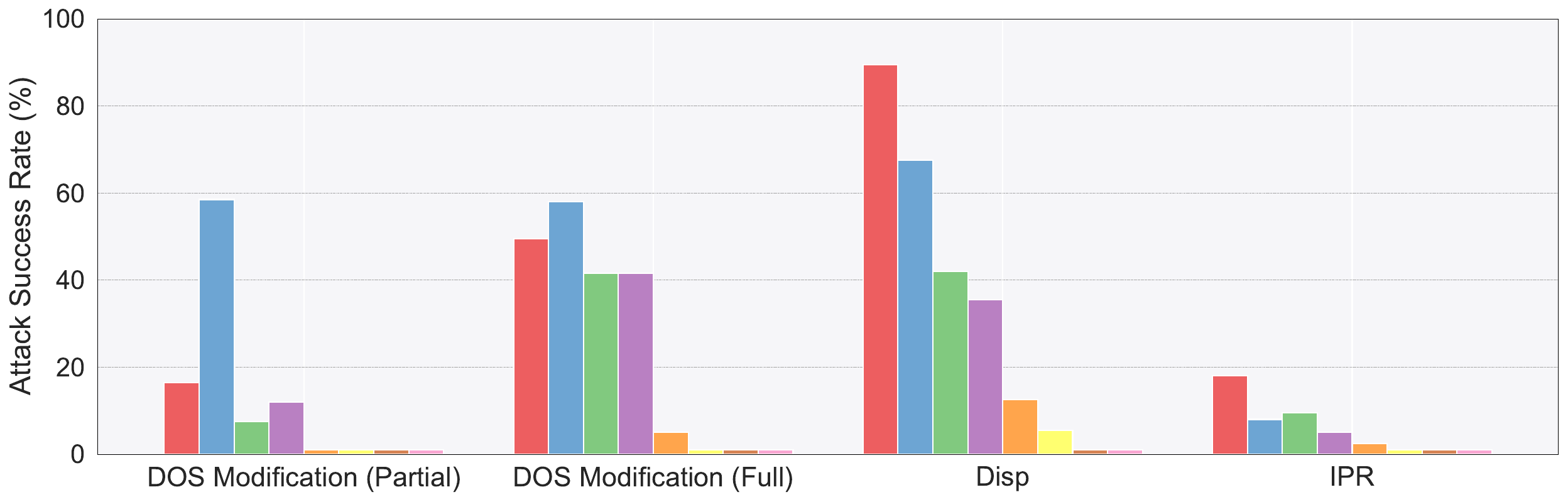}%
    }
    \caption{Attack Succes Rate (ASR) \% for white-box attacks on all models}
    \label{fig:wb_asr}
\end{figure}

Figure \ref{fig:wb_asr} shows the attack success rate (ASR) for different attacks in the white-box setting. We find that -- most attacks have less ASR on DRSM-n models than MalConv by a large margin. For example, FGSM append attack has $82.50\%$ ASR on MalConv whereas $10.0\%$ and $7.0\%$ on DRSM-4 and DRSM-8, respectively.
Moreover, for $n \geq 16$ in DRSM-n models, the ASR for all white-box attacks is $(1\%{\sim}5\%)$. 
%Moreover, ASR on DRSM-12, DRSM-16, DRSM-20, and DRSM-24 models is very marginal. 
We got the highest ASRs on MalConv model for DOS Extension ($98.00\%$) and Disp ($89.50\%$) attack, while the ASRs on DSRM-n models were in range of $(1\% {\sim} 72\%)$ and $(1\% {\sim} 42\%)$, respectively. 
%Recall from the Section \ref{sec:evaluation} that -- `$n$ in DRSM-n' has a positive correlation with certified robustness. While this pattern is still existent in empirical robustness (from Figure \ref{fig:wb_asr}), it is not as dominant as the theoretical one. 

%Interestingly, we found that -- the attacks that modify the header fields has marginally higher ASR on DRSM-8 than DRSM-4. It can be the case that -- DRSM-8 generates the ablated sequences from original file in a way that -- `perturbation in header fields' impact higher number of windows than other DRSM models. 

Though Disp and IPR attacks fall outside of our threat model, surprisingly, DRSM-n can still provide good robustness against them (Figure \ref{fig:wb_asr}). 
Here is a potential explanation: Transformed bytes by Disp and IPR at different places get divided into multiple ablated sequences and thus, they become less impactful in altering multiple predictions compared to one prediction. An interesting observation is that the attacks that modify the header fields have marginally higher ASR on DRSM-8 than on DRSM-4: Potentially, this is because for DRSM-8 the perturbed positions in header fields happen to cover more windows than other cases. 
%Finding from our primary investigation is -- when Disp and IPR attacks are run against the base classifier MalConv, they transform code at multiple different places to evade the prediction by the model. However, those transformations become less influential when they get divided into multiple ablated sequences and they have to alter multiple predictions. Thus, our DRSM-n model can provide better robustness even against such sophisticated attacks.  

\begin{figure}[tbp]
        \centering
        \includegraphics[width=1.0\linewidth]{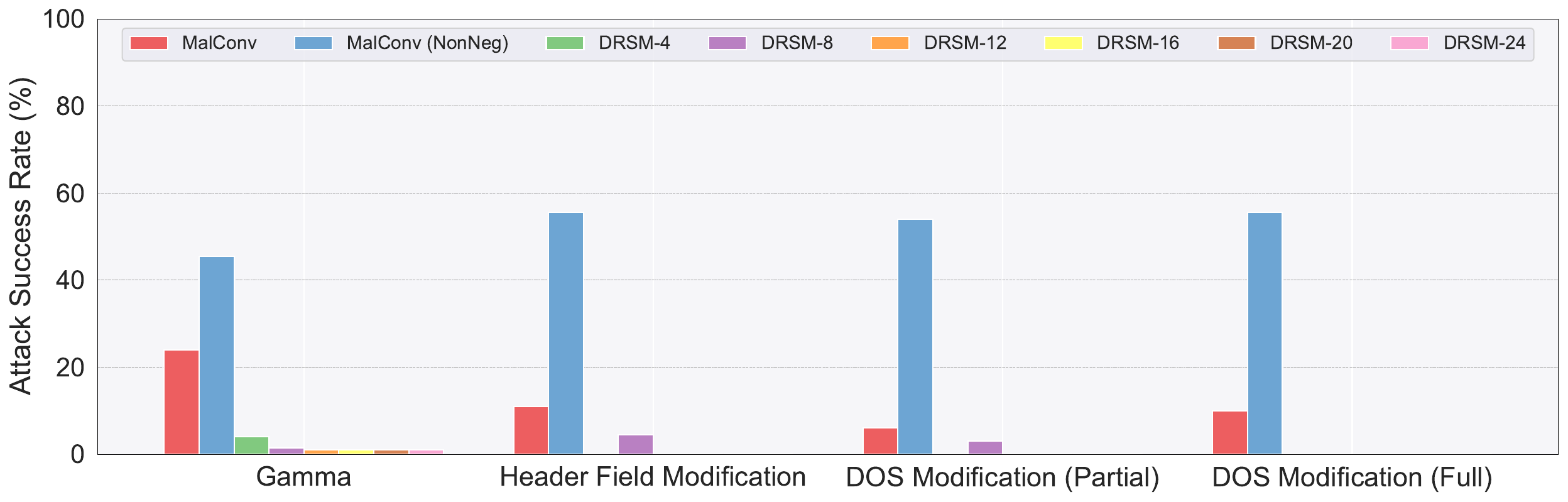}
        \caption{Attack Succes Rate (ASR) \% for black-box attacks on different models}
        \label{fig:bb_asr}
\end{figure}

We also evaluated the models against black-box attacks using genetic optimizers. For example, GAMMA attack extracts payload from benign programs and injects them into malware by querying the model. From Figure \ref{fig:bb_asr}, GAMMA has $24\%$ ASR on MalConv whereas $(4{\sim}1)\%$ on DRSM-n models. While it is true that -- these black-box attacks have less ASR on MalConv compared to the white-box ones, still DRSM-n models outperform. Interestingly, we found that MalConv(NonNeg) suffers in query-based black-box attacks, though it has been believed as a robust model for a long time. Our results are consistent with some recent works, e.g., Dropper attack by \citet{wang2022black}, MPass, GAMMA attack by \citet{wang2023mpass}, Goodware string append by \citet{ceschin2019shallow}.

%\section{Discussion}
%\input{sections/discussion}

\section{Conclusion}
\label{sec:conclusion}
In this work, we tried to find a solution for the `accuracy vs. robustness' double-edged sword in the malware field. We showed that certified defense is also possible in the executable malware domain, hoping that it will open up a new paradigm of research. Besides theory, we equally emphasized the empirical robustness of our proposed DRSM. 
Nevertheless, we must acknowledge the reality that no model can provide absolute defense, and as researchers, our intent was to raise the bar for potential attackers. 
We would like to conclude by highlighting some areas and future directions our work identifies. 
Firstly, there is room for improving the standard accuracy of DRSM by introducing an additional classification layer, albeit at the expense of challenging the fundamental non-differentiable nature of the smoothing scheme, which we chose not to explore in this study. 
Secondly, recent defenses from vision, besides de-randomized smoothing, hold promise for future exploration.
Malware detection is inherently an arms race and we hope our work can facilitate future research in developing more practical defenses with our defense and dataset.

\subsubsection*{Acknowledgments}
We are immensely grateful to Keane Lucas for training our models on VTFeed dataset, and prodviding the private implementation of Disp, IPR attack (guided) to evaluate in this paper.

\bibliography{iclr2024_conference}
\bibliographystyle{iclr2024_conference}

\newpage

\appendix
\section{Appendix}
\label{sec:appendix}

\subsection{Our Published Dataset: PACE}
\label{app:dataset}

\subsubsection{Dataset Details}

\begin{figure}[h]
        \centering
        \includegraphics[width=0.60\linewidth]{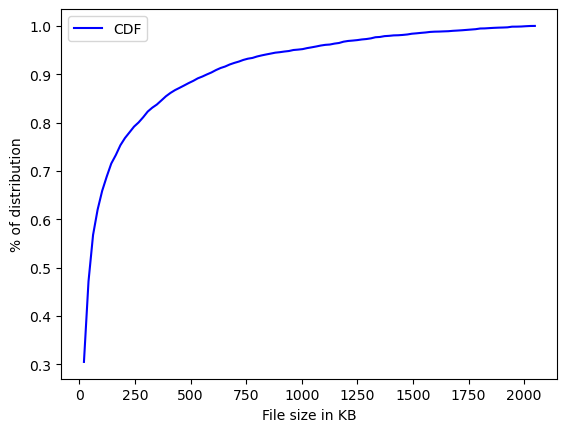}
        \caption{CDF plot of file sizes for our published benign dataset}
        \label{fig:cdf_benign_dataset}
\end{figure}

Our diverse benign dataset contains benign raw executables from $7$ different sources. Figure \ref{fig:cdf_benign_dataset} shows the cumulative distribution function (CDF) of the file sizes of our benign files. 

\textbf{Data Format. } For each benign raw executables, we are going to publish the URL link to download it from with its MD5 hash for the response (See \texttt{`dataset'} folder in our supplementary material). For example, one line from our csv file is -- 

\begin{table}[h]
  %\centering
\begin{center}
\begin{tabular}{ c|c }
    \hline
    URL & MD5 hash  \\
    \hline 
    \href{https://sourceforge.net/projects/pdfcreator/files/PDFCreator/PDFCreator%200.7/PDFCreator-Setup-0_7_0.exe/download}{\shortstack[l]{https://sourceforge.net/projects/pdfcreator/ \\ files/PDFCreator/PDFCreator\%200.7/ \\ PDFCreator-Setup-0\_7\_0.exe/download}}    & {\texttt{afaf0caffeff781f6070f2a9aeb54bdf}} \\
     \hline 
\end{tabular}
\end{center}
%\caption{Other Benign Datasets}
\label{table:other_dataset}
\end{table}

\subsubsection{Other Datasets}
\begin{table}[h]
  %\centering
  \caption{Other Benign Datasets and their public availability}
\begin{center}
\begin{tabular}{ lc }
    \hline
    Benign Datasets & Public Availability \\
    \hline 
    PACE (Our) & Raw Binary Executable \\
    Ember & only Feature Vector  \\
    VTFeed & \XSolidBrush \\
    DeepReflect & only Feature Vector \\
    BODMAS & only Feature Vector \\
     \hline 
    
      \hline
\end{tabular}
\end{center}

\label{table:other_dataset}
\end{table}

\subsubsection{Performance Degradation on Our PACE (Benign) Dataset}
\label{app:pace_degrade}
While there have been works about concept drift on malware \citep{yang2021cade, jordaney2017transcend, barbero2022transcending} and they demonstrated how malwares evolved over the time, there have been very less work on concept drift of benign files. The probable reason can be the common belief that -- benign files do not evolve or change, i.e., the distribution remains same for them. However, we evaluated a version of MalConv model on our PACE (benign) dataset that was trained on (Ember + VTFeed) dataset. 
%First type -- MalConv just trained on Ember dataset which misclassified $1.52\%$ benign files from our dataset. Second type -- 
Surprisingly, this MalConv version was misclassifying $12.22\%$ benign files from PACE dataset while it was still having $98.91\%$ test accuracy on VTFeed dataset. Recall that -- our PACE dataset is the most recent one among these (2022). It is obvious that -- these benign datasets have different distributions due to the variation in collection time. It might be the case that -- with time, different companies release (or update) their softwares for newer version of Windows, and as a result, it causes a shift in benign file distribution too. So, we would suggest researchers to report their model performance on recent datasets in future, especially when it is about security-critical domain like malware detection.

\subsection{Model Implementation}
\label{app:model_impl}

\begin{figure}[h]
        \centering
        \includegraphics[width=1.0\linewidth]{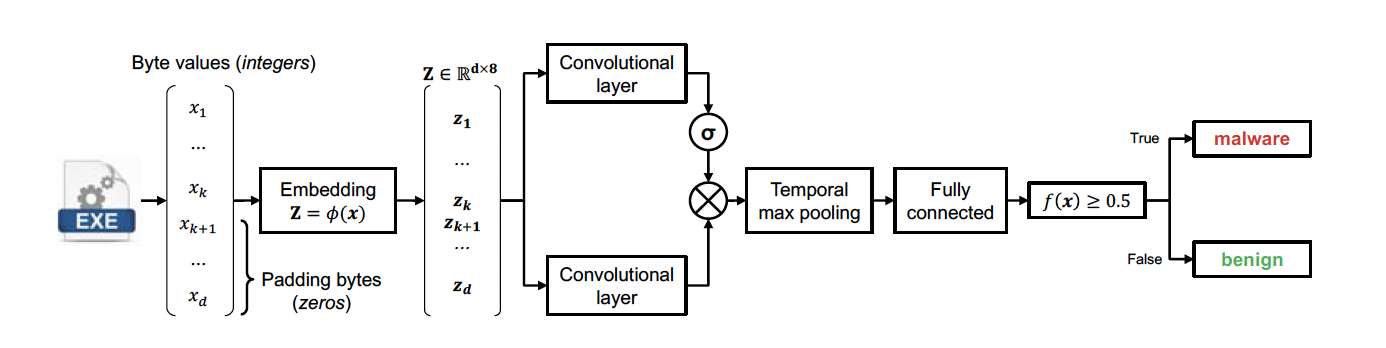}
        \caption{MalConv model architecture}
        \label{fig:malconv_model}
\end{figure}

For MalConv and MalConv (NonNeg) implementation, we used input size of $2MB$. For our optimizer, we used -- 
\begin{itemize}
    \item Optimizer: \texttt{SGD}
    \item learning-rate: $0.01$
    \item momentum: $0.9$
    \item nesterov: \texttt{True}
    \item weight-decay: $1e-3$
\end{itemize}
We used the same setting for every model -- MalConv, MalConv (NonNeg), and DRSM-n.
For training on VTFeed and our dataset, the batch size was 16 and 32, respectively. All the models were re-trained for 10 epochs. We trained the models using multiple gpus at different times. But mostly used gpus were 4 NVIDIA RTX A4000 and 2 RTX A5000.

\subsection{More on Results}
\label{app:results}

\iffalse
\smk{need to change this table}
\begin{table}[h]
  %\centering
\begin{center}
\begin{tabular}{ lccc }
    \hline
      \multirow{2}{*}{Model}  & \multicolumn{3}{c}{F1 Score $\uparrow$} \\
             & Train-set & Validation-set & Test-set \\
     \hline
    MalConv  & $99.73$ & $98.88$ & $98.61$\\
    MalConv(NonNeg) & $88.56$ & $87.56$ & $87.70$\\
    DRSM-4   & $99.49$ & $98.12$ & $98.70$\\
    DRSM-8   & $99.67$ & $97.88$ & $97.79$\\
    DRSM-12   & $96.07$ & $95.58$ & $95.98$\\
    DRSM-16   & $94.29$ & $93.00$ & $93.64$\\
    DRSM-20   & $91.17$ & $91.05$ & $91.76$\\
    DRSM-24   & $90.22$ & $89.80$ & $91.03$\\
     \hline
\end{tabular}
\end{center}
\caption{Standard Accuracy and Performance of Models}
\label{table:std_acc_app}
\end{table}
\fi

\begin{figure}[h]
        \centering
        \includegraphics[width=1.0\linewidth]{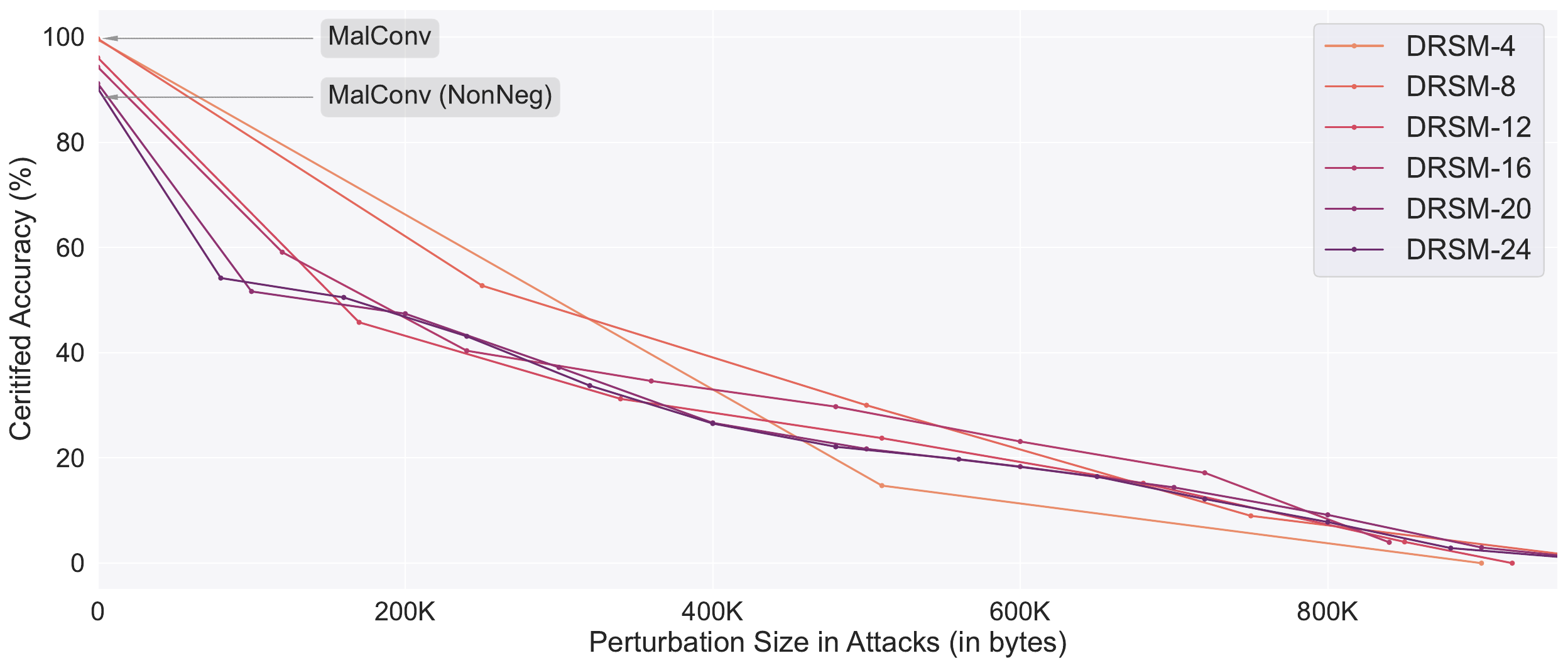}
        \caption{Certified Accuracy (\%) of DRSM-n models for different perturbation budgets (Train-set)}
        \label{fig:cert_acc_train_plot}
\end{figure}

\begin{figure}[h]
        \centering
        \includegraphics[width=1.0\linewidth]{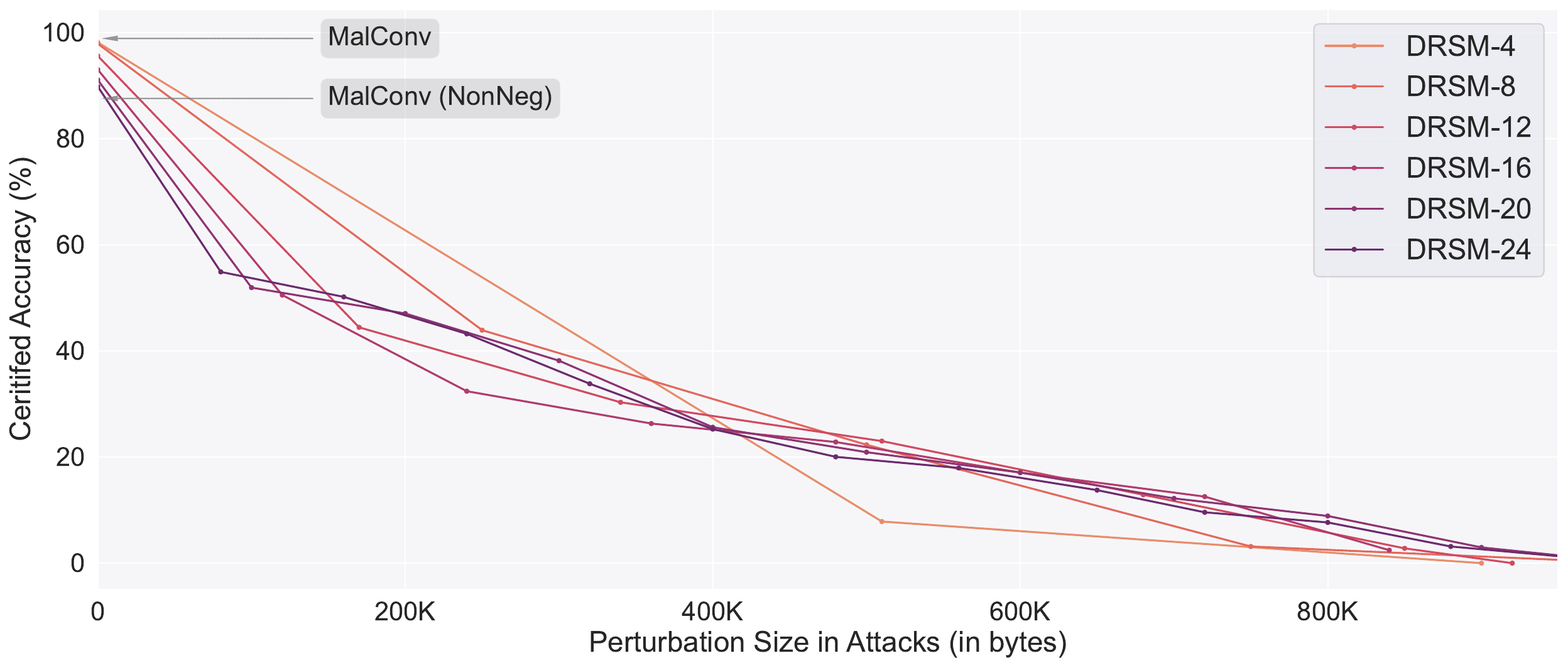}
        \caption{Certified Accuracy (\%) of DRSM-n models for different perturbation budgets (Validation-set)}
        \label{fig:cert_acc_val_plot}
\end{figure}

\begin{table}[h]
  %\centering
  \caption{Certified Accuracy (in \%) shown as a range for different $\Delta$}
\begin{center}
\begin{tabular}{ lccc }
    \hline
      \multirow{2}{*}{Model} & \multicolumn{3}{c}{Certified Accuracy (in \%) $\uparrow$} \\\cline{2-4}
      & Train-set & Validation-set & Test-set \\
     \hline
    MalConv  & $-$ & $-$ & $-$ \\
    MalConv(NonNeg) & $-$ & $-$ & $-$ \\
    DRSM-4   & $14.74$ & $7.84$ & $12.2$ \\
    DRSM-8   & $52.74 \sim 8.99$ & $43.9 \sim 3.14$ & $40.85 \sim 3.7$ \\
    DRSM-12   & $45.77 \sim 4.04$ & $44.43 \sim 2.79$ & $46.21 \sim 4.25$ \\
    DRSM-16   & $59.1 \sim 3.96$ & $50.52 \sim 2.44$ & $49.17 \sim 2.59$ \\
    DRSM-20   & $51.64 \sim 2.97$ & $51.92 \sim 2.96$ & $52.68 \sim 1.85$ \\
    DRSM-24   & $54.19 \sim 2.86$ & $54.88 \sim 3.14$ & $53.97 \sim 2.03$ \\
     \hline
\end{tabular}
\end{center}

\label{table:cert_acc_range}
\end{table}

\begin{table}[h]
\caption{Certified Accuracy (in \%) for different perturbation budget for all models}
    \centering
    
    \begin{tabular}{l c c c c}
    \hline
        \multirow{2}{*}{Model} & Perturbation & \multicolumn{3}{c}{Certified Accuracy (in \%) $\uparrow$} \\\cline{3-5}
         & Budget (in bytes) & Train-set & Validation-set & Test-set \\
    \hline
    MalConv & - & - & - & - \\
    MalConv (NonNeg) & - & - & - & - \\
    \hline
    DRSM-4 & 200K & $14.74$ & $7.84$ & $12.2$ \\
                            
    \hline
    \multirow{3}{*}{DRSM-8} & 250K & $52.74$ & $43.9$ & $40.85$ \\
                           & 500K & $30.01$ & $22.3$ & $18.85$ \\
                           & 750K & $8.99$ & $3.14$ & $3.7$ \\
    \hline
    \multirow{5}{*}{DRSM-12} & 170K & $45.77$ & $44.43$ & $46.21$ \\
                            & 340K & $31.23$ & $30.31$ & $32.9$ \\
                            & 510K & $23.76$ & $23.0$ & $24.21$ \\
                            & 680K & $15.19$ & $12.89$ & $14.05$ \\
                            & 850K & $4.04$ & $2.79$ & $4.25$ \\
    \hline
    \multirow{7}{*}{DRSM-16} & 120K & $59.1$ & $50.52$ & $49.17$ \\
                            & 240K & $40.37$ & $32.4$ & $31.79$ \\
                            & 360K & $34.62$ & $26.31$ & $23.11$ \\
                            & 480K & $29.74$ & $22.82$ & $17.01$ \\
                            & 600K & $23.12$ & $17.07$ & $11.46$ \\
                            & 720K & $17.17$ & $12.54$ & $8.32$ \\
                            & 840K & $3.96$ & $2.44$ & $2.59$ \\
    \hline
    \multirow{9}{*}{DRSM-20} & 100K & $51.64$ & $51.92$ & $52.58$ \\
                            & 200K & $47.41$ & $47.04$ & $47.13$ \\
                            & 300K & $37.2$ & $38.15$ & $38.82$ \\
                            & 400K & $26.66$ & $25.61$ & $28.47$ \\
                            & 500K & $21.71$ & $20.91$ & $21.81$ \\
                            & 600K & $18.35$ & $17.07$ & $18.48$ \\
                            & 700K & $14.39$ & $12.2$ & $13.12$ \\
                            & 800K & $9.18$ & $8.89$ & $6.65$ \\
                            & 900K & $2.97$ & $2.96$ & $1.85$ \\
    \hline
    \multirow{11}{*}{DRSM-24} & 80K & $54.19$ & $54.88$ & $53.97$ \\
                            & 160K & $50.5$ & $50.17$ & $49.91$ \\
                            & 240K & $43.12$ & $43.21$ & $45.1$ \\
                            & 320K & $33.74$ & $33.8$ & $34.75$ \\
                            & 400K & $26.54$ & $25.26$ & $26.8$ \\
                            & 480K & $22.12$ & $20.03$ & $21.26$ \\
                            & 560K & $19.76$ & $17.94$ & $17.38$ \\
                            & 650K & $16.45$ & $13.76$ & $13.68$ \\
                            & 720K & $12.19$ & $9.58$ & $8.5$ \\
                            & 800K & $7.81$ & $7.67$ & $5.36$ \\
                            & 880K & $2.86$ & $3.14$ & $2.03$ \\

    \hline
    \end{tabular}
    
    \label{table:cert_acc_all}
\end{table}

\begin{table}[h]
  \centering
  \caption{Attack Success Rate (ASR) \% of different evasion attacks in White-box setting}

\resizebox{\linewidth}{!}{
\begin{tabular}{ lcccccccc }
    \hline
        \multirow{3}{*}{Attack (White-box)}  & \multicolumn{8}{c}{Attack Success Rate (\%) $\downarrow$} \\\cline{2-9}
       & \multicolumn{7}{c}{Models} \\
             & MalConv & MalConv(NonNeg) & DRSM-4 & DRSM-8 & DRSM-12 & DRSM-16 & DRSM-20 & DRSM-24 \\
     \hline
    FGSM Append & $82.50$ & $23.5$ & $10.00$ & $7.00$ & $1.00$ & $1.00$ & $1.00$ & $1.00$ \\
    Slack Append & $79.50$ & $17.0$ & $22.50$ & $25.50$ & $1.00$ & $1.00$ & $1.00$ & $1.00$ \\
    DOS Extension & $98.00$ & $26.5$ & $72.00$ & $78.00$ & $12.50$ & $2.00$ & $1.50$ & $1.00$\\
    Header Field Modification & $26.50$ & $58.5$ & $10.00$ & $17.00$ & $1.00$ & $1.00$ & $1.00$ & $1.00$\\
    DOS Modification (Partial) & $16.50$ & $58.5$ & $7.50$ & $12.00$ & $1.00$ & $1.00$ & $1.00$ & $1.00$\\
    DOS Modification (Full) & $49.50$ & $58.00$ & $41.50$ & $41.50$ & $5.00$ & $1.00$ & $1.00$ & $1.00$\\
    Disp & $89.50$ & $67.5$ & $42.00$ & $35.50$ & $12.50$ & $5.50$ & $1.00$ & $1.00$\\
    IPR & $18.00$ & $8.00$ & $9.50$ & $5.00$ & $2.50$ & $1.00$ & $1.00$ & $1.00$\\
      \hline
\end{tabular}
}

\label{table:res_attack_wb}
%\end{sidewaystable}

\vspace{4\baselineskip}
%\begin{sidewaystable}[h]
  \centering
  \caption{Attack Success Rate (ASR) \% of different evasion attacks in Black-box setting}

\resizebox{\linewidth}{!}{
\begin{tabular}{ lcccccccc }
    \hline
        \multirow{3}{*}{Attack (Black-box)}  & \multicolumn{7}{c}{Attack Success Rate (\%) $\downarrow$} \\\cline{2-9}
       & \multicolumn{7}{c}{Models} \\
             & MalConv & MalConv(NonNeg) & DRSM-4 & DRSM-8 & DRSM-12 & DRSM-16 & DRSM-20 & DRSM-24 \\
     \hline
    GAMMA & $24.00$ & $45.50$ & $4.00$ & $1.50$ & $1.00$ & $1.00$ & $1.00$ & $1.00$ \\
    Header Field Modification & $11.00$ & $55.50$ & $0.00$ & $4.50$ & $0.00$ & $0.00$ & $0.00$ & $0.00$\\
    DOS Modification (Partial) & $6.00$ & $54.00$ & $0.00$ & $3.00$ & $0.00$ & $0.00$ & $0.00$ & $0.00$\\
    DOS Modification (Full) & $10.00$ & $55.50$ & $0.00$ & $0.00$ & $0.00$ & $0.00$ & $0.00$ & $0.00$\\
      \hline
\end{tabular}
}

\label{table:res_attack_bb}
\end{table}

%\newpage
%\subsection{Why not Randomized-Smoothing?}
%\input{appendix/random_smoothing}

\clearpage
\subsection{Attacks}
\label{app:attack_app}

\begin{figure}[h]
        \centering
        \includegraphics[width=1.0\linewidth]{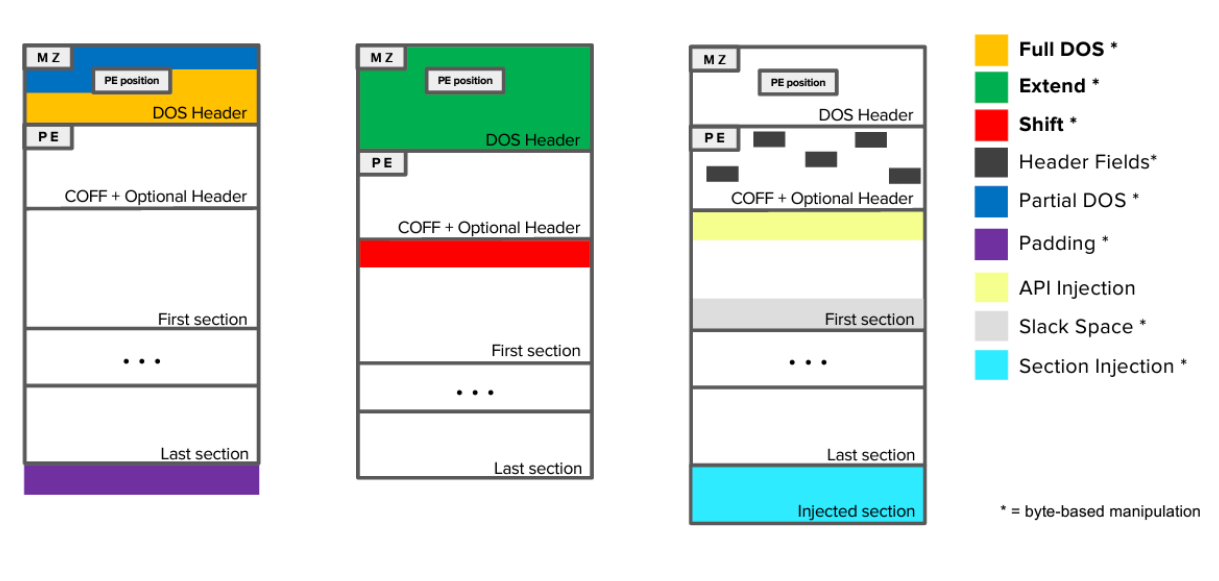}
        \caption{Graphical Representation of the locations perturbed by different attacks with adversarial payloads}
        \label{fig:attack_pe}
\end{figure}

\blfootnote{Figure \ref{fig:attack_pe} is taken from \citet{demetrio2021adversarial}.}

\subsubsection{FGSM Append Attack}
Append attack in adversarial malware was first proposed by \citet{kolosnjaji2018malware}. In this attack, authors added some noise at the end of a malware file computes by gradient of the model. However, the first proposed method was computationally heavy. Later, it was improved by \citet{kreuk2018deceiving} using Fast Gradient Signed Method (FGSM) motivated by \citet{goodfellow2014explaining}. In Figure \ref{fig:attack_pe}, the `Padding' label (purple) depicts the FGSM Padding (or Append) attack.

In our experiment, we kept a padding budget of $10$KB ($=0.5\%$ of the input file size) and ran the attack for $10$ iterations. We noticed that, for some malwares the model prediction was $1.0$ for which the attack failed.

\subsubsection{Slack Append Attack}
This attack was an incremental work on \citet{kreuk2018deceiving} by \citet{suciu2019exploring}. Unlike the previous one, the attacker can inject the payload in between of sections. The find the gaps between consecutive sections (called `slack spaces') in a binary by $RawSize - VirtualSize$, and use  that gap to inject gradient-generated adversarial bytes. Since these slack spaces can be at multiple places, this attack is partially inside our threat model. In Figure \ref{fig:attack_pe}, the `Slack Space' label (grey) indicates this attack.

In our experiment, we followed the same parameter as the previous one, and ran it for 10 iterations by keeping the padding budget $10$KB. We want to mention that -- though this attack seems more evasive than the previous one, for larger perturbation budget FGSM Append is more successful than this one. This was found in original paper, and our result is consistent with this finding too. 

\subsubsection{DOS Extension Attack}
This attack creates a new space by extending the DOS header. Attacker increases the offset to PE header and modify the file structure accordingly. In these extended spaces, attacker can put adversarial bytes to evade a model \citep{demetrio2021adversarial}. Since the extension is on a contiguous portion (header) of the file, it falls under our threat model. In Figure \ref{fig:attack_pe}, the `Extend' label (green) refers to this attack.
We ran this attack on our `attack set' for $10$ iterations with $10^{-3}$ penalty regularizer.

\subsubsection{DOS Modification Attack}
There are 2 versions of this attack -- Partial\citep{demetrio2019explaining}, and Full\citep{demetrio2021adversarial}. In DOS header, two important bytes are -- magic number \texttt{MZ} and real offset \texttt{0x3c}. The former attack modifies bytes in between of these two bytes while the latter one modified every bytes in the DOS header except those two. So, the `full' modification version is more evasive than the `partial' one. This attack is shown in blue and yellow color in Figure \ref{fig:attack_pe}. We ran this attack on our models for 10 iterations.

\subsubsection{Header Field Modification Attack}
This attack was proposed by \citet{nisi2021lost}. They analyzed the discrepancies among tools and PE file formats. Thus, they found a set of bytes (or modifications) that can potentially evade a malware classifier. Since this attack modifies bytes at multiple different places but they are constrained only in the PE header, it is partially inside our threat model. In Figure \ref{fig:attack_pe}, the `Header Fields' label (black) shows how this attack changes header fields in PE header. We ran this attack for $20$ iterations.  

\subsubsection{Disp (Code Displacement) Attack}
\begin{figure}[h]
        \centering
        \includegraphics[width=0.4\linewidth]{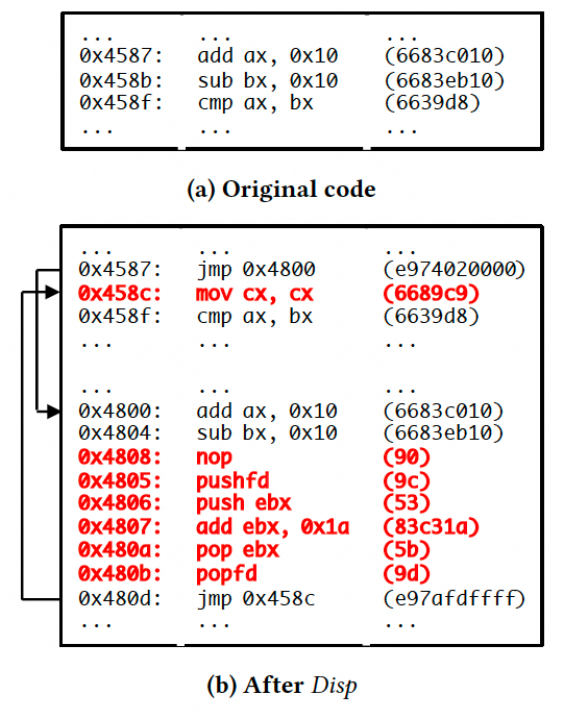}
        \caption{An example of Disp attack}
        \label{fig:disp}
\end{figure}

In this attack, the attacker has to use to disassemble a malware first. Then the attacker displace consecutive instructions in a basic block. Such displacements are usually done \texttt{jmp} and \texttt{nop} instructions. \citet{lucas2021malware} proposed this attack for the first time. Figure \ref{fig:disp} shows an example of such attack.

We collected the private implementation of the guided version of this attack from the authors. We ran Disp-1 (the perturbation budget is $1\%$ of the binary size) for 100 iterations. 

\subsubsection{IPR (In-Place Randomization) Attack}
\begin{figure}[h]
        \centering
        \includegraphics[width=1.0\linewidth]{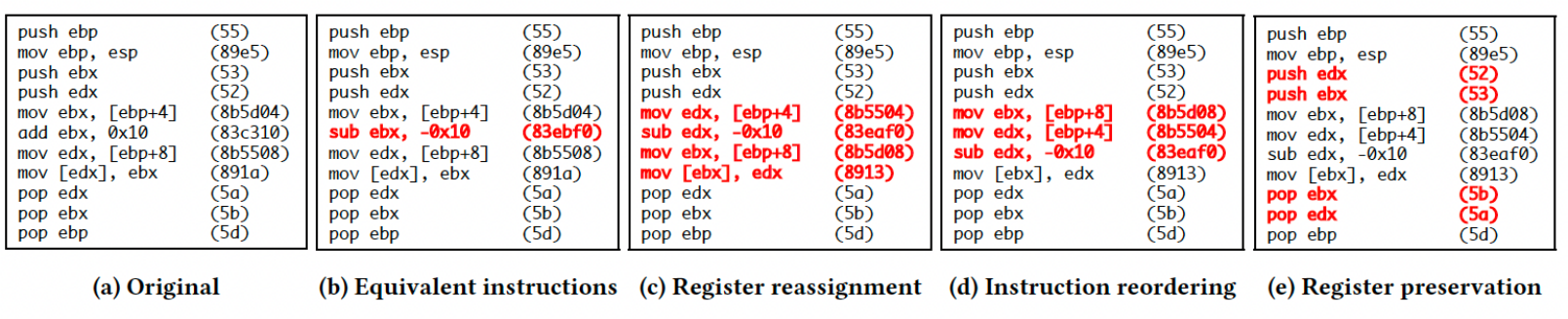}
        \caption{An illustration of IPR attack}
        \label{fig:ipr}
\end{figure}

Like the previous attack, attacker has to disassemble the malware here. Then attacker can apply four types of transformations -- 
i) replacing instructions with equivalent ones, ii) reassigning registers, iii) reordering instructions using dependency graph, and iv) altering register's push and pop order.
These transformations do not necessarily change the file size but it modifies the code at many different places. So, this attack falls outside of our threat model. Figure \ref{fig:ipr} shows the transformation types with an example. We collected the private implementation for this attack from authors of \citet{lucas2021malware}.

\subsubsection{GAMMA Attack}
\begin{figure}[h]
        \centering
        \includegraphics[width=1.0\linewidth]{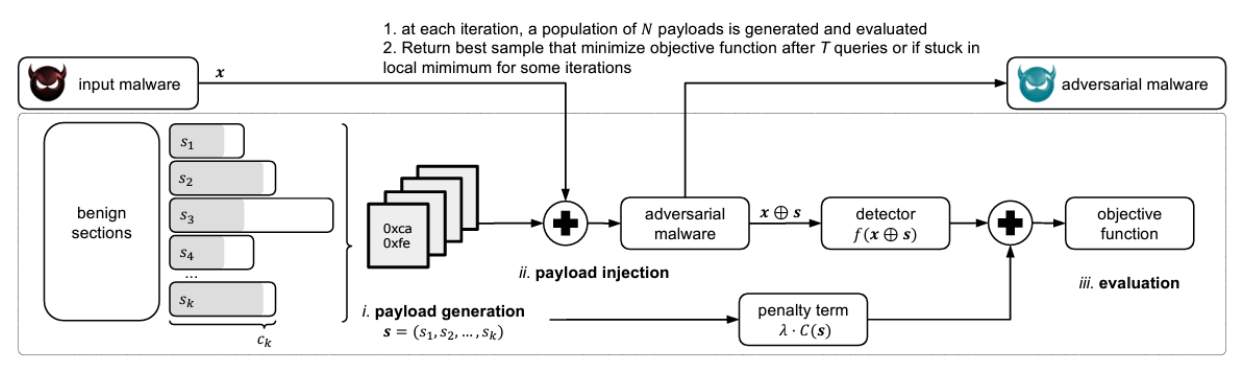}
        \caption{Overview of GAMMA attack}
        \label{fig:gamma}
\end{figure}

This attack was first proposed by \citet{demetrio2021functionality}. Though it was a common belief that -- goodware (or benign) payload (or string) can be added to a malware to evade a model, they are the first to propose a query-based black-box method for this. In this attack, the attacker generates payload from some benign programs, then inject them into a malware and return the best subset of generations by querying the model. Figure \ref{fig:gamma} shows the overview of the attack.

In our experiment, we ran the hard-label version of GAMMA attack with section injection. We set the population size and query as $200$, and ran it for $20$ iterations. For payload extraction, we used the \texttt{.data} section of benign files.

\blfootnote{For Disp and IPR attacks, we used IDAPro disassembler.}
\blfootnote{Figure \ref{fig:disp} and \ref{fig:ipr} are taken from \citet{lucas2021malware}.}
\blfootnote{Figure \ref{fig:gamma} is taken from \citet{demetrio2021functionality}.}

\end{document}